\documentstyle[12pt,epsfig,rotate]{article}
\begin{document}
\thispagestyle{empty}
\vspace*{1in}
\begin{center}
{\large\bf Investigation of inelastic $^{40}$Ca{\it (p,p')X}
reaction at 1 GeV }\\
\vspace*{1cm}
 O.V.~Miklukho (1), G.M.~Amalsky (1), V.A.~Andreev
(1), O.Ya.~Fedorov (1), K.~Hatanaka (2), D.~Ilyin (1), A.A.~Izotov
(1), A.Yu.~Kisselev (1), M.P.~Levchenko (1), T.~Noro (2),
A.N.~Prokofiev (1), R.~Revenko (1), H.~Sakaguchi (2),
A.V.~Shvedchikov (1), A.~Tatarenko (1), S.I.~Trush (1), A.A.~Zhdanov (1)\\
\vspace*{1cm}
(1) {\it St.Petersburg Nuclear Physics Institute, Gatchina, RUSSIA}\\
(2)  {\it Research Center for Nuclear Physics, Osaka, JAPAN}\\
\end{center}
\vspace*{1cm}
 The polarization of the secondary protons in the
inelastic ({\it p,p'}) reaction on the $^{40}$Ca nucleus and the
relative cross sections of this reaction at 1 GeV of the initial
proton energy were measured in a wide range of the scattered
proton momenta ({\bf K}) at lab. angles $\Theta=13.5^\circ$ and
$\Theta=21.0^\circ$. The final protons from the reaction were
detected by means of a magnetic spectrometer equipped with
multiwire - proportional chambers polarimeter.\\

{\bf Comments:} 9 pages, 3 figures, 4 tables. \\


{\bf Category:} Nuclear Experiment (nucl-ex)\\

\newpage

\begin{center}
{\bf Abstract}
\end{center}

{ The polarization of the secondary protons in the inelastic ({\it
p,p'}) reaction on the $^{40}$Ca nucleus and the relative cross
sections of this reaction at 1 GeV of the initial proton energy
were measured in a wide range of the scattered proton momenta
({\bf K}) at lab. angles $\Theta=13.5^\circ$ and
$\Theta=21.0^\circ$. The final protons from the reaction were
detected by means of a magnetic spectrometer equipped with
multiwire - proportional chamber polarimeter.

Close to the maximum of the {\it pN} quasielatic cross section
peak, which is mainly determined by the ({\it p,pN}) reactions, a
reduction of the measured polarization of the scattered protons in
comparison with predictions of the Plane Wave Impulse
Approximation (PWIA), based on parameters of the  free elastic
proton-proton and proton-neutron scatterings, was observed as  in
other similar experiments. A big value of the polarization
obtained in the range of the scattered proton momenta  is larger
than those corresponding to the {\it pN} quasielastic peak, where
a contributions from the ({\it p,pN}) reaction are suppressed, is
related possibly to a manifestation of the cluster component of
the $^{40}$Ca wave function. }

\normalsize
\section{Introduction}
\label{sect1}

 The  present work is a part of wide experimental program in the frame of which
 medium - induced modifications of nucleon - nucleon scattering amplitudes are
 studied at PNPI synchrocyclotron with 1 GeV proton beam energy[1-7].
 In the exclusive ({\it p,2p}) experiments with different nuclei the
 polarization of both secondary protons have been measured and the shell structure of
 the investigated nuclei being clearly distinguished. In these experiments, a two-arm magnetic
 spectrometer having a high energy overall resolution was used. Both arms of the
 spectrometer were equipped with multiwire proportional chamber polarimeters.

 In the present work, the high energy arm of the spectrometer was used to
  measure the polarization ({\bf P}) of the secondary protons in the inelastic (inclusive)
 $^{40}$Ca({\it p,p'}){\it X} reaction and the relative differential cross sections of the reaction
 as a function of the scattered proton momentum ({\bf K}) at relatively
small lab. angles $\Theta=13.5^\circ$ and $\Theta=21.0^\circ$.

 The results of the experiment one assumes to use for
 estimation of the integral contribution from the multiple knockout collisions
 (multi-step processes) in the momentum region of the {\it pN} quasielastic peak, where
 the single-step ({\it p,pN}) knockout reactions are proposed to be dominant.
 Such estimations, for
 instance, have been done for the reaction $^{12}$C({\it p,p'}){\it X} at 800 MeV [8].
 According to this work the contribution from multi-step processes
in the polarization of the scattered protons with momenta close to
maximum of the {\it pN} quasielastic peak is not so large.

 A large reduction of the analyzing power Ay (in experiments with polarized proton beam) and
 secondary proton polarization P in the inclusive {\it A(p,p')X} reaction in the
 momentum
 range of the {\it pN} quasielastic peak relative to the corresponding values
 for free {\it pN} scattering was called a "relativistic signature"[9] and
 a significant part of the reduction is explained in the relativistic approaches [10].

\section{Experimental results}

The measured polarization of the scattered protons {\bf P} in the
$^{40}$Ca({\it p,p'}){\it X} reaction and relative cross sections
of the reaction are presented in Fig.~1-2 (see also Tables~1-4 and
Fig.~3 in Appendix). In Fig.~1-2, the solid and dashed curves are
the result of polarization calculations in the plane wave impulse
approximation (PWIA) proposing that the mechanism  of the reaction
is a single-step ({\it p, pN}) process, and using the final (FEP)
and initial (IEP) energy prescriptions, respectively [11]. At
given scattered proton momentum {\bf K}, averaging over the
polarization in ({\it p,p'p}) and ({\it p,p'n}) scattering was
performed in the range of residual nucleus momentum  {\bf
K}$_{A-1}$ up to 200 MeV/c using the current SAID phase shift
analysis [12].

As is seen from Fig.~1-2, the measured polarization in the region
close to maximum of the {\it pN} quasielastic peak is smaller than
that predicted in the framework of the PWIA as it was observed  in
other similar experiments [8,10,13].

At the scattered proton momenta $|${\bf K}$|$ $>$ 1590 MeV/c
(Fig.~2), where the quasielastic A{\it (p,p'N)}A-1  processes are
suppressed due to the large value of the residual nucleus momentum
$|${\bf K}$_{A-1}|>$ 200 MeV/c, the other reactions can be
dominant like to the $^{40}$Ca({\it p,p'}$^4$He)$^{36}$Ar
quasielastic scattering at the momentum value $|${\bf K}$|$ about
of 1640 MeV/c. At this momentum, the value of the measured
polarization in the reaction $^{40}$Ca({\it p,p'}){\it X} is close
to that observed in the {\it p}$^4$He elastic scattering [6], If
the latter observation is not accidental, it possibly means that
the wave function of the $^{40}$Ca nucleus contains the
$^4$He-like cluster component in the kinematics under
investigation. For study in detail, it takes to perform the
polarization measurements at different angles of the scattered
proton in coincidence with a low energy particle knocked out from
the nucleus.

\begin{figure}
\centering\epsfig{file=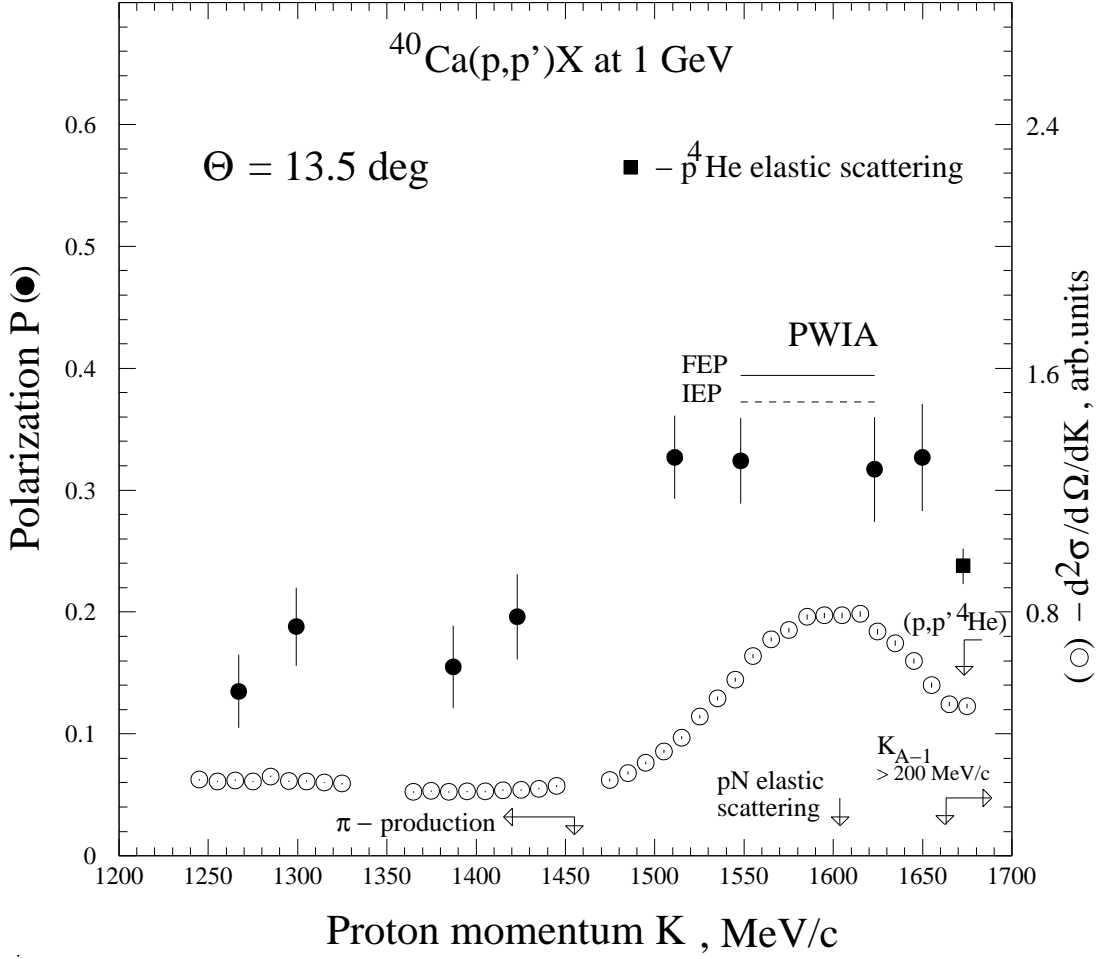,width=1.\textwidth}
\caption{Polarization {\bf P}  of the  protons scattered at angle
$\Theta=13.5^\circ$  ($\bullet$) in the inclusive reaction
$^{40}$Ca({\it p,p'}){\it X} and the relative cross section of the
reaction $\frac{d^2\sigma}{d\Omega d{\bf K}}$ - ($\circ$) as a
function of the secondary proton momentum. Solid and dashed curves
(straight lines) are the result of calculation in the PWIA using
the final (FEP) and initial (IEP) energy prescription,
respectively. The black square corresponds to the value of
polarization in the elastic {\it p}$^4$He scattering [14].}
\label{f_Four}
\end{figure}

\begin{figure}
\centering\epsfig{file=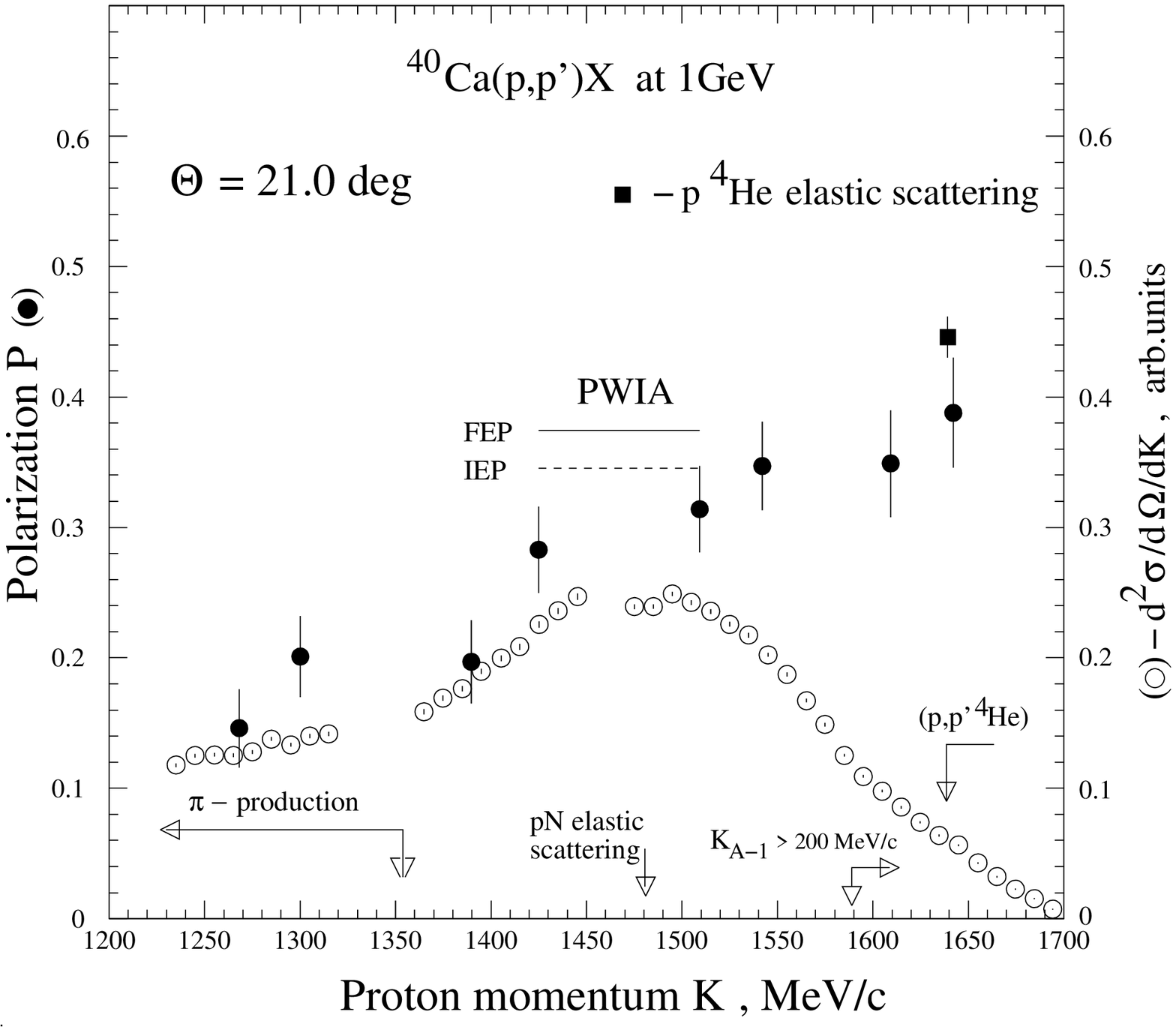,width=1.\textwidth}
\caption{Polarization {\bf P}  of the  protons scattered at angle
$\Theta=21.0^\circ$  ($\bullet$) in the inclusive reaction
$^{40}$Ca({\it p,p'}){\it X} and the relative cross section of the
reaction $\frac{d^2\sigma}{d\Omega d{\bf K}}$ - ($\circ$) as a
function of the secondary proton momentum. Solid and dashed curves
(straight lines) are the result of calculation in the PWIA using
the final (FEP) and initial (IEP) energy prescription,
respectively. The black square corresponds to the value of
polarization in the elastic {\it p}$^4$He scattering [14].}
\end{figure}

\section{Acknowledgements}

This work is partly supported by the Grant of President of the
Russian Federation for Scientific Schools, Grant-3383.2010.2.

\section{Appendix:}
\textwidth=11.5cm
\begin{center}
\small
\begin{tabular}{c|c||c|c||c|c}
\multicolumn{6}{p {\textwidth}} {\normalsize TABLE 1: The
polarization P of the scattered proton in the reaction
$^{40}$Ca({\it p,p'}){\it X}  at 1 GeV and lab. angle
$\Theta=13.5^\circ$ }\\
\multicolumn{6}{l}{ }\\
\hline \hline
K & P & K & P & K & P  \\
MeV/c &  & MeV/c &  & MeV/c &   \\
\hline 1267.1 & 0.135$\pm$0.030 & 1423.0 & 0.196$\pm$0.035 &
1623.1 & 0.317$\pm$0.043
\\
1299.3 & 0.188$\pm$0.032 & 1511.1 & 0.327$\pm$0.034 & 1649.7 &
0.327$\pm$0.044
\\
1387.3 & 0.155$\pm$0.034 & 1548.0 & 0.324$\pm$0.035 &
\\
\hline \hline
\end{tabular}
\vspace{0.5cm} \normalsize
\end{center}

\begin{center}
\small
\begin{tabular}{c|c||c|c||c|c}
\multicolumn{6}{p{\textwidth}}{\normalsize TABLE 2: The
polarization P of the scattered proton in the reaction
$^{40}$Ca({\it p,p'}){\it X}  at 1 GeV and lab. angle
$\Theta=21.0^\circ$ }\\
\multicolumn{6}{l}{ }\\
\hline
\hline
K & P & K & P & K & P  \\
MeV/c &  & MeV/c &  & MeV/c &   \\
\hline
1268.0 & 0.146$\pm$0.030 & 1424.8 & 0.283$\pm$0.033 & 1609.2 & 0.349$\pm$0.041
\\
1300.2 & 0.201$\pm$0.031 & 1509.3 & 0.314$\pm$0.033 & 1642.1 & 0.388$\pm$0.042
\\
1389.7 & 0.197$\pm$0.032 & 1542.1 & 0.347$\pm$0.034 &
\\
\hline
\hline
\end{tabular}
\vspace{0.5cm}
\normalsize
\end{center}

\begin{center}
\small
\begin{tabular}{c|c||c|c||c|c}
\multicolumn{6}{p{\textwidth}}{\normalsize TABLE 3: The relative
cross section of the reaction $^{40}$Ca({\it p,p'}){\it X}  at 1
GeV and lab. angle
$\Theta=13.5^\circ$ }\\
\multicolumn{6}{l}{ }\\
\hline
\hline
K & $\frac{d^2\sigma}{d\Omega dK}$ & K & $\frac{d^2\sigma}{d\Omega dK}$ & K & $\frac{d^2\sigma}{d\Omega dK}$  \\
MeV/c & arb.units  & MeV/c & arb.units  & MeV/c & arb.units   \\
\hline
1245.0 & 2499$\pm$26 & 1405.0 & 2111$\pm$24 & 1555.1 & 6546$\pm$65
\\
1255.1 & 2439$\pm$25 & 1415.1 & 2152$\pm$24 & 1565.2 & 7106$\pm$68
\\
1265.0 & 2476$\pm$25 & 1425.1 & 2158$\pm$25 & 1575.0 & 7403$\pm$69
\\
1274.9 & 2449$\pm$25 & 1435.1 & 2200$\pm$25 & 1585.2 & 7839$\pm$89
\\
1285.0 & 2603$\pm$25 & 1445.1 & 2291$\pm$26 & 1595.0 & 7887$\pm$89
\\
1295.1 & 2452$\pm$25 & 1474.9 & 2490$\pm$46 & 1605.0 & 7887$\pm$89
\\
1305.0 & 2446$\pm$25 & 1485.0 & 2702$\pm$42 & 1615.0 & 7935$\pm$89
\\
1314.9 & 2407$\pm$25 & 1494.9 & 3048$\pm$42 & 1624.8 & 7356$\pm$86
\\
1324.9 & 2369$\pm$25 & 1505.1 & 3420$\pm$47 & 1634.7 & 6979$\pm$84
\\
1364.9 & 2101$\pm$24 & 1515.0 & 3876$\pm$50 & 1645.0 & 6388$\pm$80
\\
1374.8 & 2127$\pm$25 & 1525.1 & 4563$\pm$55 & 1655.1 & 5602$\pm$75
\\
1384.8 & 2098$\pm$24 & 1535.1 & 5163$\pm$58 & 1665.0 & 4981$\pm$71
\\
1394.9 & 2117$\pm$24 & 1545.1 & 5782$\pm$61 & 1674.9 & 4911$\pm$70
\\
\hline
\hline
\end{tabular}
\vspace{0.5cm}
\normalsize
\end{center}

\begin{center}
\small
\begin{tabular}{c|c||c|c||c|c}
\multicolumn{6}{p{\textwidth}}{\normalsize TABLE 4: The relative
cross section of the reaction $^{40}$Ca({\it p,p'}){\it X}  at 1
GeV and angle
$\Theta=21.0^\circ$ }\\
\multicolumn{6}{l}{ }\\
\hline
\hline
K & $\frac{d^2\sigma}{d\Omega dK}$ & K & $\frac{d^2\sigma}{d\Omega dK}$ & K & $\frac{d^2\sigma}{d\Omega dK}$  \\
MeV/c & arb.units  & MeV/c & arb.units  & MeV/c & arb.units   \\
\hline
1235.1 & 1178 $\pm$14 & 1415.1 & 2085 $\pm$22 & 1574.9 & 1488 $\pm$14
\\
1245.0 & 1250 $\pm$14 & 1425.1 & 2256 $\pm$23 & 1585.1 & 1253 $\pm$11
\\
1255.1 & 1256 $\pm$14 & 1435.2 & 2359 $\pm$24 & 1595.0 & 1092 $\pm$10
\\
1265.0 & 1252 $\pm$14 & 1445.2 & 2471 $\pm$24 & 1604.9 & 978 $\pm$10
\\
1274.9 & 1278 $\pm$14 & 1475.0 & 2392 $\pm$18 & 1614.9 & 855 $\pm$9
\\
1285.0 & 1378 $\pm$15 & 1485.0 & 2392 $\pm$16 & 1624.8 & 741 $\pm$8
\\
1295.1 & 1334.$\pm$15 & 1494.8 & 2490 $\pm$18 & 1634.7 & 639 $\pm$8
\\
1305.0 & 1401 $\pm$15 & 1504.9 & 2424 $\pm$18 & 1644.9 & 564 $\pm$7
\\
1314.9 & 1417 $\pm$15 & 1514.9 & 2356 $\pm$18 & 1654.9 & 428 $\pm$6
\\
1364.9 & 1586 $\pm$19 & 1524.9 & 2258 $\pm$17 & 1664.9 & 323 $\pm$6
\\
1374.8 & 1691 $\pm$20 & 1535.0 & 2174 $\pm$17 & 1674.7 & 227 $\pm$5
\\
1385.0 & 1763 $\pm$20 & 1545.0 & 2024 $\pm$16 & 1684.6 & 156 $\pm$4
\\
1395.0 & 1898 $\pm$21 & 1554.9 & 1873 $\pm$16 & 1694.2 & 73 $\pm$3
\\
1405.3 & 1997 $\pm$22 & 1565.1 & 1672 $\pm$15 & 1704.0 & 33 $\pm$2

\\
\hline
\hline
\end{tabular}
\vspace{0.5cm} \normalsize
\end{center}

\begin{figure}
\centering\epsfig{file=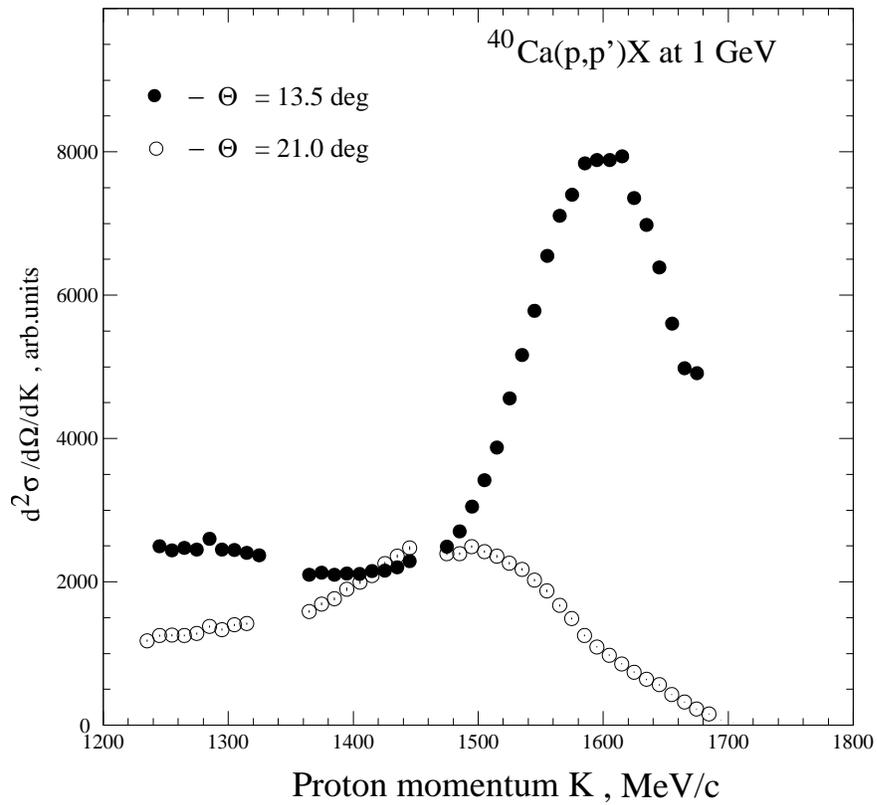,width=1.\textwidth=14cm}
\caption{The relative cross section of the inclusive reaction
$^{40}$Ca({\it p,p'}){\it X} at lab. angles $\Theta=13.5^\circ$
and $\Theta=21.0^\circ$ as a function of the scattered proton
momentum {\bf K}. } \label{f_Six}
\end{figure}

\end{document}